\documentclass[seceq]{ptptex}

\usepackage{graphicx}

\usepackage{wrapft}

\def\be{\begin{equation}}
\def\ee{\end{equation}}

\def\bi{\begin{itemize}}
\def\ei{\end{itemize}}
\def\bn{\begin{enumerate}}
\def\en{\end{enumerate}}
\def\bea{\begin{eqnarray}}
\def\eea{\end{eqnarray}}
\def\no{\nonumber}
\def\ba{\begin{array}}
\def\ea{\end{array}}
\def\bd{\begin{displaymath}}
\def\ed{\end{displaymath}}
\def\la{\langle}
\def\ra{\rangle}

\title{Ground state factorization of heterogeneous spin models in magnetic fields}

\author{
J. \textsc{Abouie}$^{1,}$\footnote{E-mail: jahan@iasbs.ac.ir}, M. \textsc{Rezai}$^{2}$, and A. \textsc{Langari}$^{2}$}

\inst{
$^1$Department of Physics, Institute for Advanced Studies in Basic Sciences, Zanjan 45137-66731, Iran\\
$^2$Department of Physics, Sharif University of Technology, Tehran 11155-9161, Iran
}

\abst{
The exact factorized ground state of a heterogeneous (ferrimagnetic) spin
model which is composed of two spins ($\rho, \sigma$) has been presented in detail. The Hamiltonian is not necessarily translational invariant and the exchange couplings can be competing antiferromagnetic and ferromagnetic arbitrarily between different sublattices to build many practical models such as dimerized and tetramerized materials and ladder compounds.
The condition to get a factorized ground state is investigated for
non-frustrated spin models in the presence of a uniform and a staggered magnetic field. According to the lattice model structure we have
categorized the spin models in two different classes and obtained their factorization conditions. The first class contains models in which their lattice structures do not provide a single uniform magnetic field to suppress the quantum correlations. Some of these models may have a factorized ground state in the presence of a uniform and a staggered magnetic field. However, in the second class there are several spin models in which their ground state could be factorized whether a staggered field is applied to
the system or not. For the latter case, in the absence of a staggered field the factorizing uniform field is unique. However, the degrees of freedom for obtaining the factorization conditions are increased by adding a staggered magnetic field.}

\begin{document}

\maketitle
\section{Introduction \label{introduction}}

Quantum nature of a spin system strongly affects its low temperature behavior and the corresponding exotic phases. The ground state is the sole candidate to explain the quantum phase transition \cite{Sach00} which takes place at zero temperature; however, the low temperature properties are highly influenced by this state.
The strongly correlated nature of the anisotropic quantum spin models prohibits us to know the ground state
exactly except in few special cases.\cite{Suzu08,Taka99}
At some particular point of the parameter space the quantum correlations vanish and the ground
state can be obtained in terms of the direct product of single particle states exactly. This
state is called a factorized state (FS) and the corresponding magnetic field is the factorizing field.
The entanglement is exactly zero at the factorizing field which corresponds to
the associated entanglement phase transition.
An exact ground state even at  particular values of the parameters gives
several information on the different properties of the model and can be implemented
to initiate a perturbation theory for more knowledge of the neighboring phases.\cite{Abou10,Langari11}

J Kurmann, et.al \cite{Kurm82} introduced the factorized ground
state of a homogeneous spin-$s$ XYZ chain in a magnetic field of arbitrary direction.
The Factorized ground state has been also investigated in the two dimensional
models through Quantum Monte Carlo simulation in terms of entanglement estimators.
\cite{Rosc04}
Recently, the dimerized XYZ spin chain in a transverse
magnetic field has been investigated which shows that the factorized point
in the parameter space of the Hamiltonian corresponds to an accidental
ground state degeneracy.\cite{Gior09}
The factorization and entanglement were also studied in Heisenberg XYZ spin arrays in a nonuniform transverse fields.\cite{Ross09}

In this article we will review our recent results \cite{Rezai10} and present the exact factorized ground state of a hetrogenuos
(ferrimagnetic) spin model which is
composed of two spins ($\rho, \sigma$) in the presence of a uniform magnetic field
on a bipartite lattice with arbitrary long
range interaction and dimensional space.
Indeed the anisotropic ferrimagnets are good candidates to study the effects of inhomogeneity and anisotropy, simultaneously.
Quantum ferrimagnets are the general class of strongly correlated magnetism
which have attracted  much interests for experimental as well as theoretical investigations.
In this study the Hamiltonian is not necessarily
translational invariant and the exchange couplings can be competing
antiferromagnetic and ferromagnetic arbitrarily between different sublattices to build
many practical models such as dimerized and tetramerized materials and ladder compounds.
In the second part of this work
we will investigate the conditions to have a factorized ground state for the non-frustrated spin models
in the presence of a uniform and a staggered magnetic field.
According to the lattice model structure we categorize the spin models in two different classes.
The first class contains models in which their lattice structures do not provide a single uniform magnetic field to suppress the quantum correlations.
Some of these models may have a factorized ground state in the presence of a uniform and a staggered magnetic field.
The ground state entanglement of these models is decreased
by adding a staggered field and vanishes at a particular values of the uniform and staggered fields.
The conditions to have a factorizing point for this type of models are strongly
dependent on their lattice structures. However, in the second class there are several
spin models in which their ground state could be factorized whether a
staggered field is applied to the system or not.
For these models, in the absence of a staggered field
the factorizing uniform field is unique. However, the degrees of freedom for
obtaining the factorization conditions are increased  by adding a staggered magnetic field.

The outline of this paper is as follows:
In section \ref{fegs} we review in details our previous results. The factorization conditions for different models which are important from experimental and theoretical points of view are presented in section \ref{examples}. The effects of a staggered field on the factorization conditions are investigated in section \ref{unst}. Finally, we summarize the results of the paper.

\section{Factorized exact ground state \label{fegs}}
In a spin model, finding an exact ground state is a nontrivial task.
However, for some special values of model parameters such as coupling constants and magnetic fields it is possible to get the exact ground state of a spin model which is called a factorized ground state ($|FS\rangle$).
The factorized ground state is a product
of single particle kets, $|\phi_i\rangle$, i.e
\be
|FS\rangle = \bigotimes_{i}|\phi_i\rangle.
\ee
At the factorizing point the quantum correlations are zero and an entanglement phase transition is occured.
In our earlier work\cite{Rezai10} we introduced the factorizing point and its corresponding ground state for
a general class of ferrimagnets in the presence of a magnetic field.
Here, we review our previous results concisely and employ them to obtain the factorizing
field of different practical spin models.
The schematic representation of the factorized ground state are also shown for these models, clearly.

The ferrimagnetic Hamiltonian in the presence of a magnetic field can be
written in the following form
\begin{equation}
\label{m.b.h}
H=\sum_{i,r}\left[ \zeta_i \hat{\zeta}_{i+r}
(J^x_r\sigma^x_{i}\rho^x_{i+r}+J^y_r\sigma^y_i\rho^y_{i+r})+
J^z_r\sigma^z_i\rho^z_{i+r}
\right]+h \sum_{i}(\sigma^z_i+\rho^z_{i}),
\end{equation}
where $i=(i_1, i_2, i_3)$ and $r=(r_1, r_2, r_3)$ are representing
the three dimensional
index on the lattice.
The ferromagnetic (F) or antiferromagnetic (AF) interactions are defined
by $\zeta_i, \hat{\zeta}_{i+r}=\pm1$.
The type of interactions can be changed differently on any short or long ranged
interacting pairs by the product sign of $\zeta_i \hat{\zeta}_{i+r}$ which is
ferromagnetic for negative and antiferromagnetic for positive sign.
A remark is in order here, the Hamiltonian in Eq.(\ref{m.b.h})
is a sum of two-site Hamiltonians (\ref{2h}) where the two spins
can be far from each other. However, the interaction between each pair of
($\sigma_i, \rho_{i+r}$) can depend on the distance, $r$ via $J^{\mu}_r$ with different strength and
also be F or AF arbitrarily imposed by $\zeta_i$ and $\hat{\zeta}_{i+r}$.

Before finding the exact ground state of the Hamiltonian (\ref{m.b.h}),
first we consider a two-site spin model which is composed of
two spins $\sigma=\frac12$ and $\rho=1$. The interaction between the two spins is a fully anisotropic Heisenberg type given by the following Hamiltonian
\be
\label{2h}
H'=J^x\sigma^x\rho^x+J^y\sigma^y\rho^y+J^z\sigma^z\rho^z+ h'(\sigma^z+\rho^z),
\ee
where $h'$ is proportional to the magnetic field.
We are looking for a
factorized state satisfying the following condition
\begin{equation}
\label{hp}
 H'|\sigma\rangle|\rho\rangle=\epsilon|\sigma\rangle|\rho\rangle,
\end{equation}
in which $|\sigma\rangle$ and $|\rho\rangle$ are the single particle states.
To find the factorized state we consider $|\sigma\rangle$ and $|\rho\rangle$ as the eigenstates of Bloch operators $\vec{\sigma} \cdot \hat{n}'$ and $\vec{\rho} \cdot \hat{n}''$, respectively.
\begin{eqnarray}
&&\no\vec{\sigma} \cdot \hat{n}'|\sigma\rangle=\frac 12|\sigma\rangle ,
~~~|\sigma\rangle=e^{i\frac{\phi}{2}}\cos(\frac{\theta}{2})|+\frac{1}{2}\rangle
+e^{-i\frac{\phi}{2}}\sin(\frac{\theta}{2})|-\frac{1}{2}\rangle , \\
&& \vec{\rho} \cdot \hat{n}''|\rho\rangle=|\rho\rangle ,
~~~|\rho\rangle=e^{-i\alpha}\cos^2(\frac{\beta}{2})|+1\rangle
+\frac{\sin\beta}{\sqrt{2}}|0\rangle
+e^{i\alpha}\sin^2(\frac{\beta}{2})|-1\rangle
\end{eqnarray}
where $|\sigma\rangle$ and $|\rho\rangle$ have been written in the basis of $\sigma^z$
and $\rho^z$ eigenstates and  $\theta, \beta$ and $\phi, \alpha$ are the
latitude and azimutal angles, respectively.

There exists a solution for Eq.(\ref{hp}) at a particular value of the magnetic field,
$h'=h'_f$, and energy $\epsilon=\epsilon_f$. The factorizing field and factorizing energy are given by:
\bea
h'_f&=&\frac{\sqrt{4J^x J^y +5J^{z^2}+ J^z C_1}}{2\sqrt{2}}, \nonumber\\
C_1&\equiv&\sqrt{8(2 J^x+ J^y)(J^x+2 J^y)+9 J^{z^2}}, \nonumber\\
\epsilon&=&\frac{J^x J^y}{2J^z}-\frac{h'^2_f}{J^z}.
\label{he}
\eea
The factorized eigenstate is also characterized by the following
parameters
\bea
\label{tetaphi}
cos\theta=\frac{1-\left|\frac {A_1}{ A_2} \right|}{1+\left|\frac {A_1}{ A_2} \right|} \;\; &,& \;\; \varphi=\arg{A_1} +\frac12 \arg{A_3}, \nonumber \\
cos\beta=\frac{|A_3|-1}{|A_3|+1} \;\; &,& \;\; \alpha=-\frac{1}{2}\arg{|A_3|},
\eea
where
\bea
A_1&=&\frac{\sqrt{2}\left(J^z+h'_f+2\epsilon\right)}{J^x+J^y}, \nonumber\\
A_2&=&\frac{\sqrt{2}\left(-J^z+3h'_f+2\epsilon\right)}{J^x-J^y},\nonumber \\
A_3&=&\frac{A_1(J^y-J^x)}{\sqrt{2}\left(J^z+3h'_f-2\epsilon\right)}.\label{A1A2A3}
\eea
The expressions (\ref{A1A2A3}) justify that $A_1$ and $A_3$ are real values
which imply that their
arguments be either $0$ or $\pi$. Thus, there are four possible choices for $\alpha$ and $\phi$ which are shown in table (\ref{table}).

\begin{wraptable}{r}{\halftext}
\caption{Different possibilities for azimutal angles and the planes of spins}
\label{table}
\begin{center}
\begin{tabular}{cccccc} \hline \hline
case &$\arg A_1$&$\arg A_3$&$\alpha$&$\phi$& Plane\\
 \hline
(I)&$0$&$0$&$0$&$0$&$xz$\\
(II)&$0$&$\pi$&$-\frac{\pi}{2}$&$\frac{\pi}{2}$&$yz$\\
(III)&$\pi$&$0$&$0$&$\pi$&$xz$\\
(IV)&$\pi$&$\pi$&$\frac{\pi}{2}$&$\frac{3\pi}{2}$&$yz$\\ \hline\hline
\end{tabular}
\end{center}
\end{wraptable}
The spins are located in the $xz$-plane for choices I and III while they have
projections only in the $yz$-plane for II and IV. Without loss of generality
we can assume that the spins are located in the $xz$-plane. In fact, the spins of $yz$-plane
will fall into the $xz$-plane by interchange of $J^x \leftrightarrow J^y$. Moreover, the
orientations $(\theta,\varphi=0)$ and $(-\theta,\varphi=\pi)$ are representing the same directions. Therefore, we only consider the case (I): $\alpha=0, \varphi=0$, as a general one, the
results for this case can describe all other possibilities by a simple substitution
mentioned above.
Accordingly, this product state is defined by
two angels ($\theta, \beta$) which show the orientations of
($\overrightarrow{\sigma}, \overrightarrow{\rho}$), respectively.

{\bf Generalization to arbitrary $\sigma$ and $\rho$}:

We generalize our approach to a system of two spins with arbitrary ($\sigma, \rho$).\cite{Kurm 82}
To find the factorized state of a general
two-site ferrimagnet we perform a rotation on $\sigma$ and $\rho$ spins such that
$\overrightarrow{\sigma}$ points in ($\theta, \varphi=0$) and $\overrightarrow{\rho}$ in ($\beta, \alpha=0$) directions.
The rotation operator is $D= D^{\sigma}(0, \theta,0)D^{\rho}(0, \beta,0)$ where
\be
D^{\rho}(0,\beta,0)=D^{\rho}(\alpha=0,\beta,\gamma=0)=D^{\rho}_{z}(\alpha)D_{y}(\beta)D_{z}(\gamma),
\nonumber
\ee
is defined in terms of Euler angles and a similar expression is also considered for
$D^{\sigma}(0,\theta,0)$.
Thus, the rotation on the Hamiltonian of the generalized two-site
ferrimagnet is given by
\begin{eqnarray}
\label{2.b.h}
D^{\dag}H'D&=&\tilde{H'}=(J^z \cos\beta \cos\theta + J^x \sin\beta \sin\theta)\sigma^{z'}\rho^{z''}\\
\nonumber
&+&(J^x \cos\beta \cos\theta + J^z \sin\beta \sin\theta)\sigma^{x'}\rho^{x''}
+J^y\sigma^{y'}\rho^{y''}\\
\nonumber&+&(-J^z \sin\beta \cos\theta + J^x \cos\beta \sin\theta)\sigma^{z'}\rho^{x''}\\
\nonumber&+&(J^x \sin\beta \cos\theta - J^z \cos\beta \sin\theta)\sigma^{x'}\rho^{z''}\\
\nonumber&+&h'(\cos\beta \rho^{z''}-\sin\beta \rho^{x''}+\cos\theta\sigma^{z'}-\sin\theta \sigma^{x'}),
\end{eqnarray}
where $x', y', z'$ represents the rotated coordinates for $\sigma$ and
$x'', y'', z''$ for $\rho$.
Then, imposing the condition to have a factorized eigenstate for the Hamiltonian of Eq.(\ref{2.b.h})
fixes the following relations for the model parameters
\begin{eqnarray}
\cos\theta&=&-\frac{h'^2_f J^y+J^x(J^{z^{2}}-J^{y^2})\rho \sigma +h'_fJ^z(J^y \rho+J^x \sigma)}{h_f^{'2} J^x+J^y(J^{z^{2}}-J^{x^2})\rho \sigma +h'_f J^z(J^x \rho+J^y \sigma)}, \nonumber\\
\cos\beta&=&-\frac{h'^2_f J^y+J^x(J^{z^{2}}-J^{y^2})\rho \sigma +h'_fJ^z(J^y \sigma+J^x \rho)}{h'^2_f J^x+J^y(J^{z^{2}}-J^{x^2})\rho \sigma +h'_f J^z(J^x \sigma+J^y \rho)}, \nonumber \\
h'_f&=& \sqrt{\frac12\big(2J^x J^y \rho \sigma+(\rho^2+\sigma^2)J^{z^2}+ J^z C_2\big)}, \nonumber\\
C_2&\equiv&\sqrt{4\rho\sigma(\rho J^x+\sigma J^y)(\sigma J^x+\rho J^y)+(\rho^2-\sigma^2)^2 J^{z^2}}, \nonumber\\
\epsilon&=&\frac{J^x J^y}{J^z}\sigma\rho-\frac{h'^2_f}{J^z}.
\label{tbhe}
\end{eqnarray}
The factorizing field $h'_f$ depends on the coupling constants $J^{x,y,z}$ and the spin magnitudes, i.e $h'_f=h'_f(J^x,J^y,J^z;\sigma,\rho)$.


{\bf Lattice Hamiltonian}

Let us utilize this procedure to attain the factorizing conditions of the generalized many-spin Hamiltonian (\ref{m.b.h}).
The factorized eigenstate for the Hamiltonian (\ref{m.b.h})
is a bi-angle state which is written in the following form
\begin{equation}
\label{fs}
|FS\rangle=\bigotimes_{i\in A_{\sigma}, j\in B_{\rho}}|\zeta_i \theta\rangle
 |\hat{\zeta}_j \beta\rangle.
\end{equation}
This state ($|FS\rangle$)
is an eigenstate of the Hamiltonian (\ref{m.b.h})
if the angle $\zeta_i \theta$ ($\hat{\zeta}_{i+r} \beta$) be consistent
with all pair of interactions originating from $\sigma_i$ ($\rho_{i+r}$) on
sublattices $A_{\sigma}$ ($B_{\rho}$).
The former condition is
satisfied if the interaction between each pair ($\sigma_i, \rho_{i+r}$) is
the same for all directions while depending on distance ($r$) as
\begin{equation}
\label{constraint}
J^{\mu}_r=\lambda(r)J^{\mu}, \;\;\; \mu=x,y,z, \;\;\; \lambda(r)>0.
\end{equation}
Under these constraints the
factorized state (Eq.(\ref{fs})) is an eigenstate of the lattice Hamiltonian with the characteristic
angles ($\theta, \beta$) defined in Eq.(\ref{tbhe}).
In other words, the necessary conditions to have a factorized eigenstate for
the lattice Hamiltonian in Eq.(\ref{m.b.h}) is that the long range
interactions satisfy the constraint given by Eq.(\ref{constraint}).
The factorizing field is given by
\be
\label{hf}
h_f= h'_f \sum_{r=0}^{N_r} \lambda(r),
\ee
where $N_r$ is the number of spins on sublattice $B_{\rho}$ ($A_{\sigma}$) which interacts with a single spin $\sigma$ ($\rho$) on sublattice $A_{\sigma}$ ($B_{\rho}$) and $h'_f$ is given by Eq.(\ref{tbhe})


We should now show that the factorized state defined in Eq.(\ref{fs}) is indeed the
ground state of the lattice Hamiltonian given by  Eq.(\ref{m.b.h}). Again, let us first
consider the special case of $\sigma=\frac12$ and $\rho=1$. The two-site spin
Hamiltonian is diagonalized exactly.  According to the obtained energy levels
the factorized state has the lowest energy.
It is the ground state of the pair
Hamiltonian at the factorizing field, $h'_f$, as far as
\be
\label{pair-condition}
J^z > \mbox{Sign}(- J^x J^y) \times \mbox{min}\{|J^x|, |J^y|\}.
\ee
The lattice Hamiltonian is the sum of pair Hamiltonians
\be
H=\sum_{i, j} H_{i,j},
\ee
which can be written as a sum of positive definite terms by
adding a constant equal to the number of interacting pairs times the
ground state energy of a single pair, i.e
\be
{\tilde H}=H- N N_r \epsilon=\sum_{i, j} (H_{i,j}-\epsilon),
\ee
where $N$ is the number of spins in each sublattice.
One can simply show that the factorized state ($|FS\rangle$, Eq.(\ref{fs}))
is an eigenstate of ${\tilde H}$ with zero energy, i.e ${\tilde H}|FS\rangle=0$.
Accepting the condition (\ref{pair-condition}), it is approved that
${\tilde H}$ has a non negative spectrum which proves that $|FS\rangle$ is
its ground state. Consequently, $|FS\rangle$ is
the ground state of $H$ with energy $N N_r \epsilon$.

Similar calculations have been also done for ($\sigma=1$, $\rho=2$),
($\sigma=\frac12$, $\rho=2$) and ($\sigma=\frac32$, $\rho=2$).
The surprising result is that in all cases the factorized state
will be the ground state if
$J^z > \mbox{Sign}(- J^x J^y) \times \mbox{min}\{|J^x|, |J^y|\}$
which is the same condition for ($\sigma=\frac12$, $\rho=1$).
We conclude that the condition (\ref{pair-condition}) is independent
of the size of spins and is general for an arbitrary magnitude of
$\sigma$ and $\rho$. Thus, the factorized state of the Hamiltonian (\ref{m.b.h})
which is indicated by (\ref{tbhe}) is an exact ground sate of the model.


{\it Magnetization and Staggered magnetization}-
One of the advantage of finding the exact ground state at the factorizing point
is to find the exact value of different order parameters such as magnetization (ferromagnetic order),
staggered magnetizations (antiferromagnetic order) and string order parameters (Haldane phase).
As a matter of fact, knowing the exact value of an order parameter at a special point of external
parameters, enables us to gain the exact phases of the system at that point.  At the factorizing
point the transverse and longitudinal components of the magnetization per unit cell are as follows
\begin{eqnarray}
&\no\left[ M_{||}\right]_{h=h_f}=\frac{1}{N}\sum_{j=1}^{N}(\sigma\cos\theta_j+\rho\cos\beta_j)=|\sigma\cos\theta+\rho\cos\beta|,\\
&\no\left[ M_{\perp}\right]_{h=h_f}=\frac{1}{N}\sum_{j=1}^{N}(\sigma\sin\theta_j+\rho\sin\beta_j)=|\sigma\zeta\sin\theta
+\rho\hat{\zeta}\sin\beta|,\\ \label{mag}
\end{eqnarray}
where   $\zeta =\frac 1N \sum{\zeta_j}$ and $\hat{\zeta} =\frac 1N\sum{\hat{\zeta}_j} $,
$\cos\theta$
and $\cos\beta$ are given by Eq. (\ref{tbhe}). Similarly, the staggered magnetization per unit cell is given by
\begin{equation}
\left[SM_{||}\right]_{h=h_f}=|\sigma\cos\theta-\rho\cos\beta|, ~~~~ \left[SM_{\perp}\right]_{h=h_f}=|\sigma\zeta\sin\theta-\rho\hat{\zeta}\sin\beta|. \label{smag}
\end{equation}
For instance, let us consider the heterogenous ferrimagnetic XXZ spin-$(\rho,\sigma)$ chain
with couplings $J^x=J^z=1$ and $J^y=\Delta$,
in the presence of a transverse magnetic field $(h^z)$. At the factorizing field,
$h_f=2(\rho^2+\sigma^2+2\Delta\sigma\rho)^{1/2}$, the magnetization
 and staggered magnetization are $\left[M_{||}=M_z\right]_{h=h_f}=\frac{h_f}{2}$,
$[M_{\perp}=M_{x,y}]_{h=h_f}=0$ and
$\left[SM_z\right]_{h=h_f}=2h_f^{-1}(\rho^2-\sigma^2)$, $\left[SM_x\right]_{h=h_f}=4\sigma\rho h_f^{-1}(1-\Delta^2)^{1/2}$
, $SM_y=0$, respectively. For the special case of $\sigma=1/2$, $\rho=1$ and $\Delta=0.5$
the magnetization is $M_z\simeq 1.32$ and staggered magnetization becomes $SM_x\simeq 0.65$.
These values are confirmed by the numerical density matrix renormalization group (DMRG)
computations \cite{Langari11}.
Moreover, the full phase diagram of the anisotropic
 ferrimagnetic spin-$(1,1/2)$ chain has been investigated by utilizing DMRG
 and three types of spin wave approximations.\cite{Langari11}.

For the homogeneous spin models, $\sigma=\rho$, $|\theta|$ is equal to $|\beta|$ and the factorized
bi-angle state reduces to a single-angle state. In this case the magnetizations and the staggered
magnetizations are  $\left[ M_{||}\right]_{h=h_f}=|\sigma\cos\theta|$, $\left[ M_{\perp}\right]_{h=h_f}=0$
and $\left[SM_{||}\right]_{h=h_f}=0$, $\left[SM_{\perp}\right]_{h=h_f}=|\sigma\sin\theta|$, respectively.
For the anisotropic antiferromagnetic spin-$1/2$ XXZ model in a transverse magnetic field defined by couplings
$J^x=J^z=1$ and $J^y=\Delta$, the factorizing field, magnetization and staggered magnetization
are $h_f=(2(1+\Delta))^{1/2}$, $\left[M_z\right]_{h=h_f}=\frac{h_f}{4}$ and
$\left[SM_x\right]_{h=h_f}=\frac 12(\frac{1-\Delta}{2})^{1/2}$, $SM_y=0$, respectively.
In the following section we bring different examples in which their ground state magnetization and
staggered magnetization at the factorizing point are given by Eqs. (\ref{mag}) and (\ref{smag})
while the angles $\theta$ and $\beta$ should be replaced from Eq.(\ref{tbhe}).


\section{Some examples\label{examples}}
The main goal of this section is to introduce the factorized ground state and its
corresponding factorizing field for some spin models which have been extensively investigated
in the literature. These models are important from experimental and theoretical points of view. The essential task is to write each lattice Hamiltonian in the general form of Eq.(\ref{m.b.h}) with couplings (\ref{constraint})
where the parameter $\lambda(r)$ defines the range of interaction of a single spin with the others.
To have a clear explanation for the factorizing
field let us emphasis again that $h'_f$ is a function of the exchange coupling constants $J^x, J^y$ and $J^z$,
in addition to the spin magnitudes
\be
h'_f= h'_f(J^x, J^y, J^z; \sigma, \rho)
\label{hf-new}
\ee
which has been given by Eq.(\ref{tbhe}).

\subsection{Homogenous Spin Chains With Bond Alternations}

The results of the previous section can be simply applied to the homogenous spin models
($\rho=\sigma=s$). In this case the interaction can be defined between the spins within the
same sublattice providing no frustrations.

\subsubsection{Spin-$s$ Chains With F-AF and AF-AF Bond Alternations}

In this part we  obtain the factorizing field of three important classes of spin models with the following Hamiltonian
\begin{equation}
H=\sum_{i=1}^N\sum_{\mu=x,y,z}(J_1^{\mu}S_{2i-1}^{\mu} S_{2i}^{\mu}+J_2^{\mu}S_{2i}^{\mu}S_{2i+1}^{\mu})+h\sum_i^N S_i^z
\end{equation}
where $J_1^{\mu}=J_{AF}^{\mu,1}>0$ and $J_2^{\mu}$ is $J_F^{\mu}<0$ or $J^{\mu,2}_{AF}>0$ for the antiferromagnetic-ferromagnetic (AF-F) and bond alternating antiferromagnetic (${\rm AF}_1-{\rm AF}_2$) models, respectively. These models have a rich field induced quantum phases such as Luttinger liquid phase\cite{Saka95,Abou08},
which has been attracted many attentions and motivates us to study their factorized ground state.


\begin{itemize}

\item {\it Antiferromagnetic-Ferromagnetic spin-1/2 chains}:

The copper compound chains such as copper nitrate, copper bromide and copper chloride are the
spin-1/2 Heisenberg chain with AF-F bond alternation.
In these systems each spin interacts with two others via ferromagnetic and antiferromagnetic couplings
as shown in Fig.\ref{FAF}.
Thus the parameter $\lambda(r)$ takes two values, 1 and $\Lambda=|\frac{J_F^{\mu}}{J_{AF}^{\mu}}|$.
The factorizing field is then $h_f=(1+\Lambda)h_f^{\prime}$ where $h'_f= h'_f(J^x_{AF}, J^y_{AF}, J^z_{AF}; \sigma=\rho=1/2)$.

\item {\it Antiferromagnetic-Ferromagnetic spin-1 chain}:

Ni$_2$(EDTA)$\cdot$H$_2$O is the bond alternating Antiferro-Ferro spin-1 chain with dominant
antiferromagnetic interaction, $\Lambda=0.9$\cite{Coro 86}. In this system, as before each spin interacts with two others via Ferromagnetic and
Antiferromagnetic couplings. Thus the parameter $\lambda(r)$ takes two values $1$ and $\Lambda$, and
$h_f=(1+\Lambda)h'_f$ where $h'_f= h'_f(J^x_{AF}, J^y_{AF}, J^z_{AF}; \sigma=\rho=1)$.

\item {\it Antiferromagnetic-Antiferromagnetic spin-1/2 chain}:

The compound (CH$_3$)$_2$CHNH$_3$Cu(Cl$_x$Br$_{1-x}$)$_3$ is effectively represented by
 AF-AF or AF-F spin chain for different concentration, $x$. At $x=0$ or low concentration
the model is AF-AF Heisenberg chain while for $x=1$ it behaves as AF-F one.\cite{Naka 03}
The ground state of this system is also factorized at $h_f=(1+\Lambda)h'_f$ where $h'_f= h'_f(J^x_{AF}, J^y_{AF}, J^z_{AF}; \sigma=\rho=1/2)$
and $\Lambda=|\frac{\tilde{J}^{\mu}_{AF}}{J^{\mu}_{AF}}|$ for low concentrations or
 $\Lambda=|\frac{J^{\mu}_F}{J^{\mu}_{AF}}|$ for $x=1$.

\begin{figure}[h]
\centerline{\includegraphics[width=8cm]{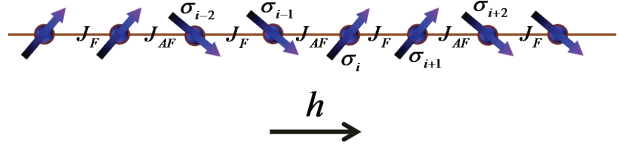}}
\caption{Antiferro-ferromagnetic bond alternating spin ($\sigma=s=\rho$) chain. The horizontal arrow denoted by $h$ shows the direction of
the magnetic field.}
\label{FAF}
\end{figure}
\end{itemize}

\subsubsection{Nersesyan-Luthur model}

Consider a bond alternating spin-1/2 XYZ chain with the following Hamiltonian
\be
H=\sum_i^N \sum_{\mu=x,y,z} J^{\mu}(1+(-1)^i\delta^{\mu})S_i^{\mu}S_{i+1}^{\mu}+h\sum_i^NS_i^z,
\ee
where $\delta^{\mu}$
is the bond alternation parameter.
Using a nonlocal unitary transformation, one can map this model to the two coupled quantum Ising  chains\cite{Ners94} which is related to the 2D Ashkin-Teller model\cite{Kohm81}.
The N-color Ashkin-Teller model was introduced to investigate some aspects related to the non-universal critical behavior presented by the usual 2D Ashkin-Teller model\cite{Gres81}. We are interested in
getting the factorizing field and factorized ground state of this model.
Based on the constraint of our method (Eq.(\ref{constraint})),
we have to consider  only $\delta^{\mu}=\delta$. Thus, the factorizing field is given by
\bea
h_f=h_f^{\prime}[(1+\delta)+(1-\delta)]=2h_f^{\prime}, \nonumber \\
h'_f= h'_f(J^x, J^y, J^z; \sigma=\rho=S)
\eea
It is worth to mention that the factorizing field of this model
is equal to the corresponding one of the uniform XYZ and $\textit{spin-Peierls}$ models.

\subsection{Ladder geometry}
\begin{figure}[h]
\centerline{\includegraphics[width=9cm]{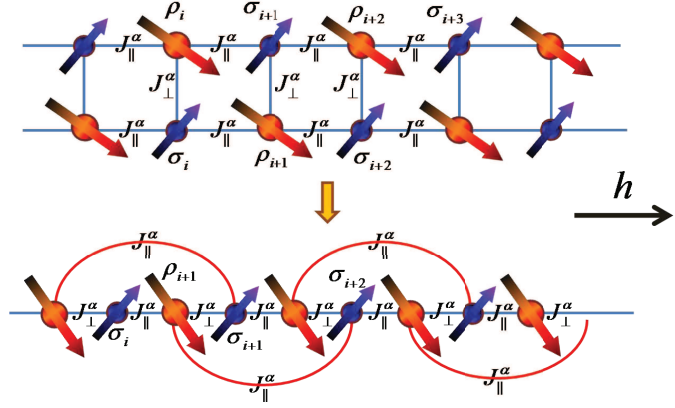}}
\caption{Ferrimagnets on ladder geometry. (top) Ladder view, (bottom) it is seen as
a chain with long range interactions. The arrows show the configuration of spins in the
factorized ground state. The horizontal arrow denoted by $h$ shows the direction of
the magnetic field.}
\label{ladder}
\end{figure}

As another example consider a ferrimagnetic model on ladder geometry\cite{Dago96} which has been
shown in Fig.\ref{ladder}-top. The Hamiltonian of a two-leg ladder is given by
\bea
\label{ladd} H&=&\sum_{i=1}^{N}\sum_{\alpha=x,y,z}
\Big(J_{\perp}^{\alpha}\sigma_{i}^{\alpha}\rho_{i}^{\alpha}
+J_{||}^{\alpha}\sigma_{i}^{\alpha}(\rho_{i-1}^{\alpha}+
\rho_{i+1}^{\alpha})\Big)+h\sum_{i=1}^N(\sigma_i^z+\rho_i^z),
\eea
where $N$
is the number of spins in each sublattic and $J_{||}$, $J_{\perp}$ are
the exchange couplings along the legs and rungs, respectively. The model represents a
ferrimagnetic ladder with antiferromagnetic ($J_{||}>0$) or ferromagnetic ($J_{||}<0$) legs in a magnetic field ($h$).
This Hamiltonian can also be regarded as a chain with long range interactions Fig.\ref{ladder}-bottom.
As it is clear from Fig.\ref{ladder} each spin $\sigma(\rho)$ interacts with three other spins of $\rho(\sigma)$. Thus, $\lambda(r)$ represents these three terms of interactions.
We choose $\lambda=1$ for the interactions along legs and
$\tilde{\Lambda}=|\frac{J_{\perp}^{\alpha}}{J_{||}^{\alpha}}|$
for the rung ones.
According to the constraint which is defined in Eq.(\ref{constraint}) the parameter $\tilde{\Lambda}$
should be independent of $\alpha$-direction.
Finally, we get the factorizing field as follows:
\be
h_f=h'_f (2+\tilde{\Lambda}) \;\;;\;\;
h'_f=h'_f(J_{||}^x, J_{||}^y, J_{||}^z; \sigma, \rho).
\ee
The factorized ground state is depicted schematically  by the arrows for
the case $J_{\perp}, J_{||}>0$
in Fig.\ref{ladder}.

\subsection{Anisotropic Triangular Lattices}

\begin{figure}[h]
\centerline{\includegraphics[width=8cm]{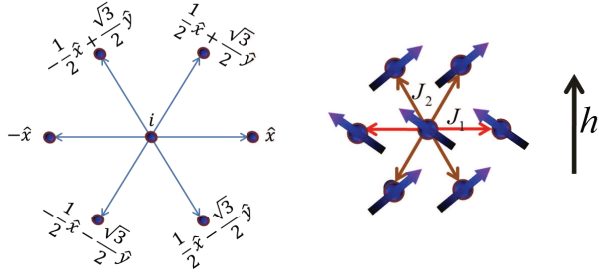}}
\caption{Triangular lattice and the corresponding factorized ground state.}
\label{TAL}
\end{figure}

The triangular lattice Hamiltonian
in the presence of a magnetic field is as follows
\begin{equation}
{\cal H}=\sum_{\la i,j\ra}\sum_{\mu=x,y,z}
\frac{J_{ij}^{\mu}}{2}S_i^{\mu}S_j^{\mu}+h\sum_iS_i^y,
\end{equation}
where $\langle\dots\rangle$ denotes nearest neighbor, $j=i+{\bold r}$ and ${\bold r}$'s indicate
the position vectors of $S_i$'s six nearest neighbors (See Fig.\ref{TAL}). We can write these vectors in the cartezian coordinate as $\pm\hat{x}$ and $\pm\frac{1}{2}\hat{x}\pm\frac{\sqrt{3}}{2}\hat{y}$, where $\hat{x}$
and $\hat{y}$ are unit vectors. Let us consider a model with the following interactions: $J_{i,i\pm\hat{x}}=J_1<0$ and $J_{i,i\pm\frac 12\hat{x}\pm\frac{\sqrt{3}}{2}\hat{y}}=J_2>0$. This model describes the properties of a non-frustrated triangular system. Thus, $\lambda(r)$ takes six values, two values are $\Lambda=|\frac{J^{\mu}_F}{J^{\mu}_{AF}}|$ and the four others are $1$. We get
\begin{equation}
h_f=(4+2\Lambda)h_f^{\prime}\;\;;\;\;h'_f= h'_f(J^x_{2}, J^y_{2}, J^z_{2}; \sigma=\rho=s).
\end{equation}

\subsection{Anisotropic Honeycomb Lattices}
\begin{figure}[h]
\centerline{\includegraphics[width=5cm]{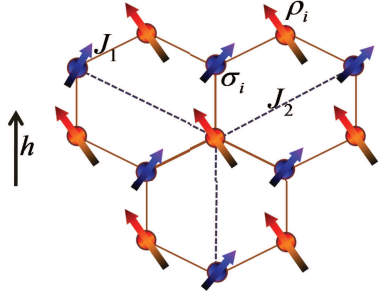}}
\caption{schematic representation of the factorized ground state of a honeycomb lattice.}
\label{HCL}
\end{figure}
Let us consider the following Hamiltonian
\be
H=\sum_{i,r}\sum_{\mu=x,y,z}J_{r}^{\mu}\sigma_i^{\mu}\rho_{i+r}^{\mu}+
h\sum_i(\sigma_i^y+\rho_i^y),
\ee
where $J_r^{\mu}$'s are coupling constants which are shown in Fig.\ref{HCL}. For the considered $J_1$-$J_2$ model, vectors $r$ in the cartezin coordinate are as follows:
\bea
\nonumber&& \hat{y},~~ -\frac{1}{2}\hat{y}+\frac{\sqrt{3}}{2}\hat{x},~~ -\frac{1}{2}\hat{y}-\frac{\sqrt{3}}{2}\hat{x},~~ -2\hat{y},~~ \hat{y}+\sqrt{3}\hat{x},~~ \hat{y}-
\sqrt{3}\hat{x}
\eea
Indeed, each spin ($\rho$) in one sublattice interacts with six spins ($\sigma$) of another sublattice and parameter $\lambda(r)$ takes six values. Three of them are $1$ and the others are $\Lambda=|\frac{J^{\mu}_2}{J^{\mu}_1}|$. Thus the factorizing field is as follow
\be
h_f=3h'_f(1+\Lambda),
\ee
where $h'_f=h'_f(J^x_1, J^y_1, J^z_1; \sigma,\rho)$.

\section{Factorizing uniform and staggered fields}\label{unst}

In this section we will investigate the factorization conditions for the spin models
in the presence of both uniform and staggered magnetic fields.
According to the model lattice structures we categorize them in the following two different classes.

The first class contains the spin models which do not meet the constraint (\ref{constraint})
and prohibit us to get a factorized ground state as mentioned in the previous section.
Indeed the lattice structure of these models does not provide a single uniform magnetic field to suppress the quantum fluctuations.
However, this does not mean that we can not find a factorized state for
these models.
Some of them may have a factorized ground state in the presence of a uniform and
a staggered magnetic fields ($H_h=\sum_i(h_u+(-1)^ih_s)S^{\mu}_i$).
The ground state entanglement of these models is decreased
by both uniform and staggered fields and vanishes at a particular values of them.
The conditions to have a factorizing point for this type of models are strongly
dependent on their lattice structures. We study this type of problems in the subsection {\it structure dependent factorization}.

In the second class, there are several spin models which satisfy the constraint
(\ref{constraint}) and their ground state could be factorized whether a
staggered field is applied to the system or not.  For these models, in the absence of a staggered field
the factorizing uniform field is unique. However, the degree of freedom is increased for
obtaining the factorization conditions by adding a staggered magnetic field.
In other words the uniform and staggered fields which provide a product state are not unique.
We investigate the factorization conditions of such models in the subsection {\it structure independent factorization}.

\subsection{Structure Dependent Factorization}

Here, we mention some spin models which do not have a factorized ground state only in
the presence of a uniform magnetic field.
Indeed, the structure of the lattice Hamiltonian impels us to add a staggered field
for attaining a product ground state.
\subsubsection{Ferrimagnets with Tetramerization}

\begin{figure}[h]
\centerline{\includegraphics[width=8cm]{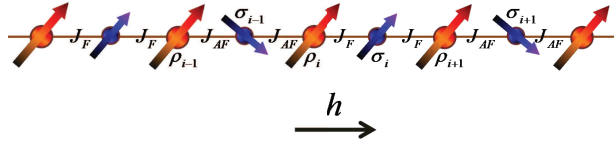}}
\caption{Schematic representation of the factorized ground state of a tetrameric ferrimagnetic chain.}
\label{TBA}
\end{figure}

Let us consider a ferrimagnetic spin chain with trimerized and tetramerized bond alternations.
Quantum ferrimagnets can be realized by
bond alternating spin-$s$ chains with a novel fashion of bonding.
For instance the spin-1/2 chain with F-AF-AF bond alternation behaves as a spin ($1,1/2$) ferrimagnetic chain at low temperatures.
Another example is Cu(3-clpy)$_2$(N$_3$)$_2$(3-clpy=3-chloropyridine) which is a real tetrameric
F-F-AF-AF spin-1/2 chain. The interaction is of the following Heisenberg type:
\bea H=\sum_iJ_F^{\mu}\sigma_{2i}^{\mu}\left(\rho_{2i-1}^{\mu}+\rho_{2i}^{\mu}\right)+
J_{AF}^{\mu}\sigma_{2i+1}^{\mu}\left(\rho_{2i}^{\mu}+\rho_{2i+1}^{\mu}\right).
\label{t.b.a}
\eea
In Fig.\ref{TBA} we have shown that each $\rho$ spin interacts with two $\sigma$ spins
ferromagnetically and antiferromagnetically.
However, the sequence of F and AF interactions are not the same for all $\rho$ spins.
Thus the parameter $\lambda$ depends on both displacement vectors, $r$ and the spins
position vectors $i$. In the other words $\lambda=\lambda(i,r)$.
Therefore, we can not choose a unique set of parameters for $\lambda$ to find the
factorizing field by making use of the method described in the previous sections.
Adding a staggered field reduces
the entanglement of the ground state and creates a product factorized ground state which is shown in Fig. (\ref{TBA}), schematically.
The factorized ground state is attained at a particular value of the uniform and the staggered fields.

For finding a condition to get a factorized ground state let us add the following terms to the Hamiltonian (\ref{t.b.a}):
\be
\sum_i\left(\left[h_u+(-1)^ih_s^{\sigma}\right]\sigma_i^z
+\left[h_u+(-1)^ih_s^{\rho}\right]\rho_i^z\right),
\ee
where $h_u$ is the uniform magnetic field, $h_s^{\sigma}$ and $h_s^{\rho}$ are proportional
to the staggered fields applying to $\sigma$ and $\rho$, respectively. Employing this
procedure, the factorizing field is given by the following parameters
\be
h_u=(1+k)h'_f,~~~~~h_s^{\sigma}=(1-k)h'_f,~~~~~h_s^{\rho}=0.
\ee
where $k=|\frac{J_{AF}^{\mu}}{J_F^{\mu}}|$ and $h'_f=h'_f(J_F^x, J_F^y, J_F^z; \sigma,\rho)$ given by Eq.(\ref{tbhe}).
The factorized ground state is gained when both uniform and staggered fields are applied only
on one type of spins, $\sigma$,
whereas the magnetic field on $\rho_i$'s is a uniform one.

\subsubsection{Nearest antiferromagnetic and next nearest ferromagnetic interactions}

\begin{figure}[h]
\centerline{\includegraphics[width=8cm]{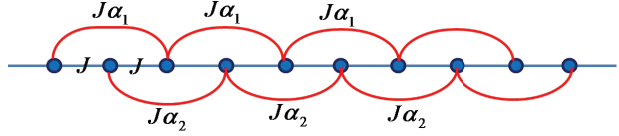}}
\caption{A lattice which does not meet the constraint (\ref{constraint}) since
the next nearest neighbor ferromagnetic interaction has two different values ($J^{\mu} \alpha_1 \neq J^{\mu} \alpha_2$).}
\label{J1J2}
\end{figure}

The second example which violates the constraint (\ref{constraint}) is shown in Fig.\ref{J1J2}.
If we apply the field $h_{\alpha_1}~(h_{\alpha_2})$ to the spins on
bond $J^{\mu}\alpha_1<0~(J^{\mu}\alpha_2<0)$ or equivalently apply
the uniform field $h_u=\frac12(h_{\alpha_1}+h_{\alpha_2})$ and the staggered
field $h_s=\frac12(-1)^i(h_{\alpha_1}-h_{\alpha_2})$ to all spins, the ground
state of the system can experience a factorized form. In fact, the quantum correlations
are decreased and goes to zero at the following values of magnetic fields:
\be
h_{\alpha_1}=2h'_f(1+\alpha_1),~~~~~~h_{\alpha_2}=2h'_f(1+\alpha_2),
\ee
where $h'_f=h'_f(J^x,J^y,J^z; \sigma=\rho=s)$ given in Eq. (\ref{tbhe}).

\subsection{Structure Independent Factorization}\label{sec-us}

In this subsection we obtain the conditions of factorization for a general
ferrimagnetic systems in the presence of a uniform and a staggered field.
The ferrimagnetic Hamiltonian can be written as
\begin{eqnarray}
\label{m.b.h-us}
H_{us}&=&\sum_{i,r}\left[ \zeta_i \hat{\zeta}_{i+r}
(J^x_r\sigma^x_{i}\rho^x_{i+r}+J^y_r\sigma^y_i\rho^y_{i+r})+
J^z_r\sigma^z_i\rho^z_{i+r}
\right] \nonumber \\
&+&(h_u+h_s)\sum_{i}\sigma^z_i+(h_u-h_s)\sum_{i}\rho^z_{i},
\end{eqnarray}
where $h_u$, $h_s$ are proportional to the uniform and staggered fields and all the other
parameters are the same as in Eq.(\ref{m.b.h}).
To obtain the factorized ground state of the Hamiltonian (\ref{m.b.h-us}),  we work on the two spin model and employ the results to the many body Hamiltonian.
For this purpose, let us consider the following two-spin Hamiltonian
\be
\label{2h-us}
\hat{H}'=J^x\sigma^x\rho^x+J^y\sigma^y\rho^y+J^z\sigma^z\rho^z+ (h'_u+h'_s)\sigma^z+(h'_u-h'_s)\rho^z.
\ee
Similar to our previous procedure (section \ref{fegs}) we perform a rotation on $\sigma$ and $\rho$
spins such that
$\overrightarrow{\sigma}$ points in
($\theta, \varphi=0$) and $\overrightarrow{\rho}$ in ($\beta, \alpha=0$) directions resulting
to the following rotated Hamiltonian
\begin{eqnarray}
D^{\dag}\hat{H}'D&=&\left(J^z \cos\beta \cos\theta + J^x \sin\beta \sin\theta\right)\sigma^{z'}\rho^{z''}\nonumber \\
&+&\left(J^x \cos\beta \cos\theta + J^z \sin\beta \sin\theta\right)\sigma^{x'}\rho^{x''}
+J^y\sigma^{y'}\rho^{y''} \nonumber \\
&+&\left(J^x \cos\beta \sin\theta-J^z \sin\beta \cos\theta\right)\sigma^{z'}\rho^{x''} \nonumber\\
&+&\left(J^x \sin\beta \cos\theta - J^z \cos\beta \sin\theta\right)\sigma^{x'}\rho^{z''} \nonumber\\
&+& h'_{\rho}\left(\cos\beta \rho^{z''}-\sin\beta \rho^{x''}\right)
+h'_{\sigma}\left(\cos\theta\sigma^{z'}-\sin\theta \sigma^{x'}\right),
\label{2.b.h-us}
\end{eqnarray}
where $x', y', z'$ represents the rotated coordinates for $\sigma$ and
$x'', y'', z''$ for $\rho$. The new magnetic fields are defined as $h'_{\sigma}=h'_u+h'_s$ and $h'_{\rho}=h'_u-h'_s$.
Imposing the condition to have a factorized eigenstate for the Hamiltonian of Eq.(\ref{2.b.h-us})
fixes the following relations for the model parameters
\bea
\cos\theta&=&-\frac{h'^2_{\sigma} J^y+J^x\left(J^{z^{2}}-J^{y^2}\right)\rho \tilde{\sigma} +h'_{\sigma}J^z(J^y \rho+J^x \tilde{\sigma})}{h'^2_{\sigma} J^x+J^y\left(J^{z^{2}}-J^{x^2}\right)\rho \tilde{\sigma} +h'_{\sigma} J^z(J^x \rho+J^y \tilde{\sigma})}, \nonumber\\
\cos\beta&=&-\frac{h'^2_{\sigma} J^y+J^x\left(J^{z^{2}}-J^{y^2}\right)\rho \tilde{\sigma} +h'_{\sigma}J^z(J^y \tilde{\sigma}+J^x \rho)}{h'^2_{\sigma} J^x+J^y\left(J^{z^{2}}-J^{x^2}\right)\rho \tilde{\sigma} +h'_{\sigma} J^z(J^x \tilde{\sigma}+J^y \rho)}, \nonumber \\
h'_{\sigma}&=& \sqrt{\frac12\big(2J^x J^y \rho \tilde{\sigma}+(\rho^2+\tilde{\sigma}^2)J^{z^2}+ J^z C_2\big)}, \nonumber\\
h'_{\rho}&=&\omega h'_{\sigma}, \nonumber\\
C_2&\equiv&\sqrt{4\rho\tilde{\sigma}(\rho J^x+\tilde{\sigma} J^y)(\tilde{\sigma} J^x+\rho J^y)+(\rho^2-\tilde{\sigma}^2)^2 J^{z^2}}, \nonumber\\
\epsilon&=&\frac{J^x J^y}{J^z}\tilde{\sigma}\rho-\frac{h'^2_{\sigma}}{J^z}.
\label{tbhe-us}
\eea
where $\tilde{\sigma}=\frac{\sigma}{\omega}$ and $\omega=\frac{h'_{\rho}}{h'_{\sigma}}$ is
the ratio of two magnetic fields. The factorizing fields depend on the coupling constants $J^{x,y,z}$, the spin magnitudes and the ratio of two magnetic fields.
At zero staggered field, the parameter $\omega$ is $1$ then $\tilde{\sigma}$ is $\sigma$ and we retrieve the results of Ref. \cite{Rezai 10}, exactly.

The factorized eigenstate for the Hamiltonian of Eq.(\ref{m.b.h-us})
can be written in the following form
\be
\label{fs-us}
|FS\rangle=\bigotimes_{i\in A_{\sigma}, j\in B_{\rho}}|\zeta_i\theta\rangle
 |\hat{\zeta}_{i+r} \beta\rangle
\ee
which is the bi-angle state.
In other words, the angle $\zeta_i \theta$ ($\hat{\zeta}_{i+r} \beta$) should be consistent
with all pairs of interactions originating from $\sigma_i$ ($\rho_{i+r}$) on
sublattices $A_{\sigma}$ ($B_{\rho}$) and the interaction between
each pair ($\sigma_i, \rho_{i+r}$) satisfy the constraint Eq.(\ref{constraint}).
With the mentioned conditions the
factorized state (Eq.(\ref{fs-us})) is an eigenstate of $H_{us}$ with the characteristic
angles ($\theta, \beta$) defined in Eq.(\ref{tbhe-us}).
The factorizing field
for $H_{us}$ is given by
\be
\label{hf-us}
h_{\sigma}= h'_{\sigma} \sum_{r=0}^{N_r} \lambda(r),
\ee
The factorizing uniform and staggered fields are written in terms of $h_{\sigma}$ as follows:
\be
h_u=\frac{h_{\sigma}}{2}(1+\omega),~~~~~h_s=\frac{h_{\sigma}}{2}(1-\omega).
\ee

It is remarkable that the factorized state given by Eq. (\ref{fs-us}) is the ground
state if $|J^y_r|>J^x_r$ [see Ref.\cite{Ross 09}].

\section{Ineligible example}

Let us point to the another example which does not meet the constraint (\ref{constraint}) and is defined by the following Hamiltonian
\be
\nonumber H=\sum_{i}(1-(-1)^i\gamma)S_i^xS_{i+1}^x+(1+(-1)^i\gamma)S_i^yS_{i+1}^y
+h\sum_i^NS_i^z,
\ee
where the nearest neighbor interaction changes its value alternatively defined by $\gamma$.
This model is different from the previous cases, because
we can not find a pair of uniform and staggered field to factorize the ground state of the system.
Such models can be studied in future investigations.

\section{Summary\label{summary}}
In this paper we have investigated the factorization conditions for different practical spin models.
In the first part of our manuscript we have introduced the factorized ground state and
factorizing field of a general class of ferrimagnets in the presence of an external magnetic field.
In our context, a ferrimagnet is composed of two different spins
on a bipartite lattice with arbitrary long range interactions constrained by Eq.(\ref{constraint})
and on arbitrary dimensional space. Employing our general result, we have attained
the factorizing point of different practical spin models such as one dimensional
bond alternating spin models, spin-Peierls chains, spin ladder systems,
and two dimensional triangular and honeycomb lattices. More examples,
such as three dimensional face centered cubic (fcc) lattice in which we only consider the
bi-partite interactions between spins can be mentioned.
In the second part of our manuscript we have studied the factorization condition of the
ferrimagnetic models
in the presence of both uniform and staggered magnetic fields.
Indeed, some of the non-frustrated spin models
do not meet the constraint (\ref{constraint})
and prohibit us to get a factorized ground state.
The lattice structure of these models does not provide a single uniform magnetic field to
suppress the quantum correlations.
The ground state of these models is factorized by adding a staggered magnetic field.
The uniform and the staggered field suppress the quantum correlations and lead to a product ground state.

We have discussed the conditions to get a factorized ground state in the presence of
a staggered field and pointed out that the staggered field increases the
degrees of freedom for finding the factorizing point.

Finally, we have presented an example which does not meet our constraint (\ref{constraint})
and is not factorized by both uniform and staggered fields.

\end{document}